# Deep Learning Potential of Mean Force between Polymer Grafted Nanoparticles

Sachin M B Gautham and Tarak K Patra[*]

Department of Chemical Engineering, Center for Atomistic Modeling and Materials Design and Center for Carbon Capture Utilization and Storage, Indian Institute of Technology Madras, Chennai TN 600036, India

**ABSTRACT**:

Grafting polymer chains on nanoparticles' surfaces is a well-known route to control their self-assembly and distribution in a polymer matrix. A wide variety of self-assembled structures are achieved by changing the grafting patterns on an individual nanoparticle's surface. However, accurate estimation of the effective potential of mean force between a pair of grafted nanoparticles that determines their assembly and distribution in a polymer matrix is an outstanding challenge in nanoscience. Here, we propose a new deep learning method that learns the interaction between a pair of grafted nanoparticles from the molecular dynamics trajectory of a cluster of polymer-grafted nanoparticles. Subsequently, we carry out the deep learning potential of mean force-based molecular simulation that predicts the self-assembly of a large number of polymer-grafted nanoparticles into various anisotropic superstructures, including percolating networks and bilayers depending on nanoparticles' concentration in three-dimension. The deep learning potential of mean force-predicted self-assembled superstructures are consistent with the actual superstructures of polymer grafted nanoparticles. This deep learning framework is very generic and can accelerate the characterization and prediction of the self-assembly and phase behaviour of polymer-grafted and un-functionalized nanoparticles in free space or a polymer matrix.

Mixtures of nanoscopic objects and macromolecules are of interest to many areas of science and technology. The structure, morphology, and phase behaviour of these multicomponent systems such as colloidal suspension, composite microgels, protein crowding in a cell, and polymer composites are determined by the complex interplay between the packing entropy of nanoparticles with varying size and shape and interparticle interaction with varying strength and range.[1–8] In many of these systems, nonadsorbing macromolecules induce attractive forces between nanoparticles (NPs), which are known as depletion forces.[9–11] The depletion force has been exploited extensively to assemble nanoparticles into a variety of superstructures. Often, non-adsorbing polymer chains are grafted on spherical NPs' surfaces to direct their assembly into various complex superstructures, including sheets, rings, icosahedra, and tetrahedra.[12–14] The grafting length and graft density are two important parameters that determine the nature of a superstructure.[15,16] From one-dimensional (1D) string to three-dimensional (3D) network-like aggregates are reported to form based on these two controlling parameters.[17] On the other hand, stronger NP-polymer interaction leads to steric stabilization, dispersion, and bridging of NPs in a polymer matrix. Microscopic liquid state theory[18–23] and molecular simulations[24–26] have been successfully used to estimate the potential of mean force (PMF) between a pair of NPs that are dissolved in a homopolymer matrix. However, the complexity of this potential energy surface enhances due to the anisotropy in shape and interaction of the NPs. An accurate estimation of the PMF that governs the self-assembly, dispersion, and bridging of non-spherical nanoparticles,[27–29] nanoparticles with tethered polymers,[30,31] nanoparticles with physical roughness[32,33] are challenging. Moreover, the architecture, polydispersity, and monomer sequence of polymers – grafted chains or bare chains enhance the complexity of the PMF.[34–39] The PMF is traditionally estimated from the radial distribution function (RDF) of NPs $\left(g_{NP-NP}(r)\right)$ in a polymer matrix as $PMF = -k_B T \ln[g_{NP-NP}(r)]$, where $k_B$ and $T$ are the Boltzmann constant and temperature of the system, respectively. However, the points of grafting on an individual NP's surface are not spatially isotropic for low grafting density, and this spatially asymmetric polymer distribution causes the effective, two-body inter-NP potential to have a strong orientational dependence that produces anisotropic self-assemblies. To coarse-grain out the information regarding tether polymers, it is, therefore, desirable to replace the information lost by an effective interaction between the centroids of the NPs that captures its radial as well as orientational dependencies.[40] Therefore, the most commonly used expression of PMF tends to fail in capturing this angular-dependent effective interaction between NPs that determines their aggregation accurately.[41] Moreover, estimating the PMF of a nanocomposite system requires experimentally or computationally measured distribution of NPs in the system. Therefore, predicting the PMF in a polymer nanocomposite system a-priori is challenging.

In order to address these problems, here, we postulate that the Behler-Parrinello symmetry functions[42,43] within a deep learning framework can capture the local environment of interacting polymer grafted NPs and provide a numerically accurate approach for constructing the PMF. As a proof of concept, we consider one polymer chain tethered NP in an implicit solvent condition and establish a deep learning framework to learn and predict the PMF between them. Previous studies suggest that NPs with a single tethered chains form a wide range of complex structures from wormy micelles to hexagonally packed cylinders to gyroid to lamellar bilayers depending on the volume fraction.[44,45] Here we aim to estimate the PMF of these self-assembled structures via deep learning. We conduct coarse-grained molecular dynamics simulations (CGMD) of a small cluster of single polymer chain grafted nanoparticles (GNPs) for a range of temperatures. The configurations and the corresponding cohesive energies of these clusters that are collected across the temperature range are used to build a generalized deep neural network PMF. The central idea of this approach is to represent the total energy ($E$) of a GNP cluster as a sum of the contributions from the individual building blocks as schematically shown in Figure 1. The total energy of an aggregate of GNPs can thus be written as $E = \sum_{i=1}^{N}(E_i + \sum_{j=1}^{M} e_j)$. Here, $N$ is the total number of GNPs in an aggregate, and $M$ is the number of grafting points in a GNP. The interaction energy of the centroid of an NP and a point of grafting on the surface of an NP are $E_i$ and $e_j$, respectively. The $E_i$ and $e_j$ depend on the local chemical environment of the system and decay radially. We choose to truncate them at a predefined cut-off distance. We represent the energy surfaces of the centroid and the point of grafting using a set of Behler-Parrinello-type symmetry functions. We expect that this higher dimensional representation of the potential energy surface will capture the angular dependency of the effective interaction between NPs due to the presence of grafted polymers. Two deep neural network models (DNNs) are built to represent these two energy surfaces of the system using the GNP cluster data. We posit that these two DNNs together provide the potential of mean force between the building blocks. The trained models make accurate predictions of the energy of unseen GNP clusters. Further, we conduct large-scale MD simulations of NPs that are interacted via the deep learning potential of mean force (DL-PDF) and predict superstructures. We find that these superstructures are identical to that of a single polymer grafted nanoparticles' actual self-assembled structures. The DL-PMF is therefore able to capture the anisotropic interaction among the grafted NPs and yields accurate self-assembled superstructures. This framework is very generic and extensible to capture the PMF between a pair of filler nanoparticles in any composite system. We expect that this deep learning framework for PMF will accelerate the characterization, understanding, and prediction of microstructures and phase behavior of polymer nanocomposites and other blends.

To train the DNNs, we conduct CGMD simulations of clusters of 2 to 10 GNPs and generate the training data. Polymer chains are represented as a coarse-grained bead-spring model of Kremer and Grest[46] wherein a pair of monomers is interacted via the Lennard-Jones (LJ) potential of the form $V(r) = 4\varepsilon\left[\left(\frac{\sigma}{r}\right)^{12} - \left(\frac{\sigma}{r}\right)^{6}\right]$. The $\epsilon$ is the unit of pair interaction energy and σ is the size of a monomer. In addition, two adjacent coarse-grained monomers of a polymer chain are connected by the Finitely Extensible Nonlinear Elastic (FENE) potential of the form $E = -\frac{1}{2}KR_0^2 \ln\left[1 - \left(\frac{r}{R_0}\right)^2\right]$, where $K = 30\epsilon/\sigma^2$ and $R_0 = 1.5\sigma$ for bond length $r \leq R_0$ and $E = \infty$ for $r > R_0$. The LJ interaction between a pair of monomers is truncated and shifted to zero at a cut-off distance $r_c = 2^{1/6}\sigma$ to represent soft repulsion between them. The NP is also modeled via the LJ potential. The diameter of an NP ($D$) is $3\sigma$. The NP-NP interaction is truncated and shifted to zero at a distance $r_c = 2.5 \times 3\sigma$ to represent attraction between them. One of the end monomers of a polymer chain is affixed to an NP surface. An NP and the grafted monomer of the affixed-polymer chain move as a rigid body during an MD simulation.[47] The polymer-NP interaction is considered to be repulsive to model the nonadsorbing polymers. The polymer-NP interaction is truncated and shifted to zero at a cut-off distance $r_c = 2 \times 2^{1/6}\sigma$. We conduct implicit solvent molecular dynamics simulations of a cluster of GNP. The initial configuration of GNPs is placed in a large simulation box to form a cluster. We use the velocity-Verlet algorithm with a timestep of $0.001\tau$ to integrate the equation of motion. Here, $\tau = \sigma\sqrt{m/\epsilon}$ is the unit of time, and $m$ is the mass of a monomer, All the simulations are conducted for a range of reduced temperature $T^* = T\,k_B/\epsilon$, which is maintained by the Langevin thermostat within the LAMMPS simulation environment.[48] The temperature is varied from T*=1.5 to T*=0.2 with a step size of $\Delta T^* = 0.1$. At each temperature, the MD simulations is conducted for $10^6$ steps. During these MD simulations, we collect ~8000 configurations of GNPs clusters across the range of temperature and cluster size.

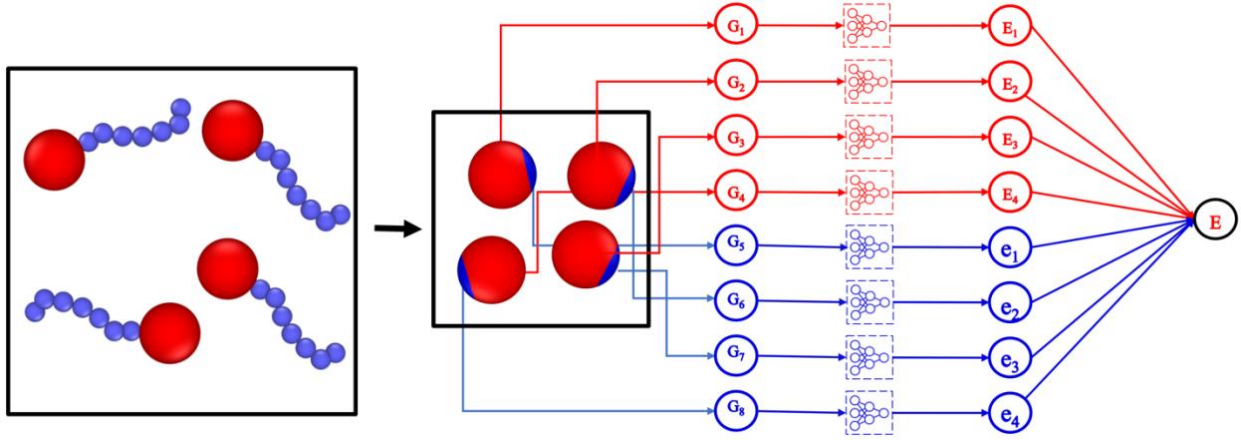

Figure 1: Structure of the high dimensional deep learning model development framework for potential of mean force between a pair of polymer-grafted nanoparticles. Within this deep learning framework, a polymer grafted nanoparticle is mapped to a patchy particle wherein the grafted polymer chain is replaced by a patch at the point of grafting. The centroid of a nanoparticle and the point of grafting are represented by two types of deep neural networks. The individual energies of the centroids of nanoparticles and the patches in a cluster are added and equated to the total energy of the cluster that is calculated using CGMD simulation.

We construct the potential energy surface using a combination of radial and angular symmetry functions. These radial and angular symmetry functions are represented as $G_i^1 = \sum_{j \neq i} e^{-\eta(R_{ij}-R_s)^2} \cdot f_c(R_{ij})$ and $G_i^2 = 2^{1-\zeta} \sum_{j,k \neq i}^{N_n} (1 + \lambda \cos\theta_{ijk})^\zeta \cdot e^{-\eta(R_{ij}^2+R_{ik}^2+R_{jk}^2)} \cdot f_c(R_{ij}) \cdot f_c(R_{ik}) \cdot f_c(R_{jk})$, respectively, where $f_c(R_{ij}) = 0.5\left[\cos\left(\frac{\pi R_{ij}}{R_c}\right) + 1\right]$ for $R_{ij} < R_c$ and $f_c(R_{ij}) = 0.0$ otherwise. The indices $j$ and $k$ run over all the neighboring particles $N_n$ within a cut-off radius of $R_c = 7.5\sigma$. To capture the local environment of a GNP accurately, we consider 25 radial symmetry functions $G^1$, each with a distinct value of $\eta$, and 25 angular symmetry functions $G^2$, each with a distinct set of $\eta, \zeta, \lambda$ values. The parameters of these 25 radial symmetry functions and 25 angular symmetry functions are reported in Table I and Table II, respectively. These symmetry functions are translationally and rotationally invariant. These sets of symmetry functions represent the energy surface of the centroid of an NP as well as the point of grafting. We choose these sets of symmetry functions based on our preliminary studies to improve the performance of the models. The variation of G¹ and G² for the selected parameter sets can be seen in supporting information (SI). A DNN architecture consists of four layers of neurons, all the neurons/nodes of a layer are connected to all the nodes in the next layer by weights in the manner of an acyclic graph. We consider two intermediate layers (hidden layers) consisting of 15 nodes each. The input layer has 50 nodes that hold 50 symmetry functions and represent the potential energy surface (PES) of a point. The output layer consists of one node that represents the potential energy of the point. Within this network topology, the three-dimensional Cartesian coordinates of a GNP are mapped into rotational and translational invariant coordinates as $G^1$ and $G^2$ symmetry functions. A cluster of *N* GNPs is represented by *2N* DNNs during the training, as schematically shown in Figure 1. We note that the architecture of all the DNNs that represent the centroid of an NP as well as the point of grafting are identical. During the training, the symmetry functions of each point are fed to the corresponding DNN via its input layer, as schematically shown in Figure 1. In every DNN, all the compute nodes in the hidden layers receive the weighted signals from all the nodes of its previous layer and feed them forward to all the nodes of the next layer via an activation function as $x_{ij} = f\left(\sum_k W_{k,j}^i x_{i-1,k}\right)$. Here, $f(x) = \tanh(x)$ is used as the activation function of all the compute nodes. As schematically shown in Figure 1, the sum of all the outputs from all the DNNs serves as the predicted energy of a GNP cluster. During the training of DNNs, all the weights are optimized to reduce the difference between the predicted energy and actual energy of a GNP cluster. We note that two types of DNNs are built during the training. One type represents the PES of the centroid of an NP, and the 2nd type represents the PES of the point of grafting on an NP's surface. They have a distinct set of hyperparameters i.e., the weights between connecting nodes. The Atomic Energy Network (AENet) software package[49] is used to build these two DNN models.

| G¹ | $\eta$ ($\sigma^{-2}$) | G¹ | $\eta$ ($\sigma^{-2}$) | G¹ | $\eta$ ($\sigma^{-2}$) | G¹ | $\eta$ ($\sigma^{-2}$) | G¹ | $\eta$ ($\sigma^{-2}$) |
|---|---|---|---|---|---|---|---|---|---|
| 1 | 0.00417 | 6 | 0.01551 | 11 | 0.0576 | 16 | 0.21386 | 21 | 0.79406 |
| 2 | 0.00543 | 7 | 0.02016 | 12 | 0.07488 | 17 | 0.27802 | 22 | 1.03229 |
| 3 | 0.00706 | 8 | 0.02621 | 13 | 0.09734 | 18 | 0.36143 | 23 | 1.34197 |
| 4 | 0.00917 | 9 | 0.03408 | 14 | 0.12654 | 19 | 0.46986 | 24 | 1.74456 |
| 5 | 0.01193 | 10 | 0.0443 | 15 | 0.1645 | 20 | 0.61082 | 25 | 2.2679 |

*Table 1: The parameter sets for the radial symmetry functions that represent the high dimensional potential energy surface for the centroid as well as the point of grafting in a polymer grafted nanoparticle.*

| G² | $\eta$ ($\sigma^{-2}$) | $\lambda$ | $\zeta$ | G² | $\eta$ ($\sigma^{-2}$) | $\lambda$ | $\zeta$ | G² | $\eta$ ($\sigma^{-2}$) | $\lambda$ | $\zeta$ | G² | $\eta$ ($\sigma^{-2}$) | $\lambda$ | $\zeta$ |
|---|---|---|---|---|---|---|---|---|---|---|---|---|---|---|---|
| 1 | 0.0004 | 1 | 2 | 8 | 0.0354 | 1 | 3 | 15 | 0.0704 | 1 | 4 | 22 | 0.1054 | -1 | 5 |
| 2 | 0.0054 | 1 | 2 | 9 | 0.0404 | 1 | 3 | 16 | 0.0754 | -1 | 4 | 23 | 0.1104 | -1 | 5 |
| 3 | 0.0104 | 1 | 2 | 10 | 0.0454 | -1 | 3 | 17 | 0.0804 | -1 | 4 | 24 | 0.1154 | -1 | 5 |
| 4 | 0.0154 | -1 | 2 | 11 | 0.0504 | -1 | 3 | 18 | 0.0854 | -1 | 4 | 25 | 0.1204 | 1 | 6 |
| 5 | 0.0204 | -1 | 2 | 12 | 0.0554 | -1 | 3 | 19 | 0.0904 | 1 | 5 | | | | |
| 6 | 0.0254 | -1 | 2 | 13 | 0.0604 | 1 | 4 | 20 | 0.0954 | 1 | 5 | | | | |
| 7 | 0.0304 | 1 | 3 | 14 | 0.0654 | 1 | 4 | 21 | 0.1004 | 1 | 5 | | | | |

*Table 2: The parameter set for the angular symmetry functions that represent the high dimensional potential energy surface of the centroid as well as the point of grafting in a polymer grafted nanoparticle.*

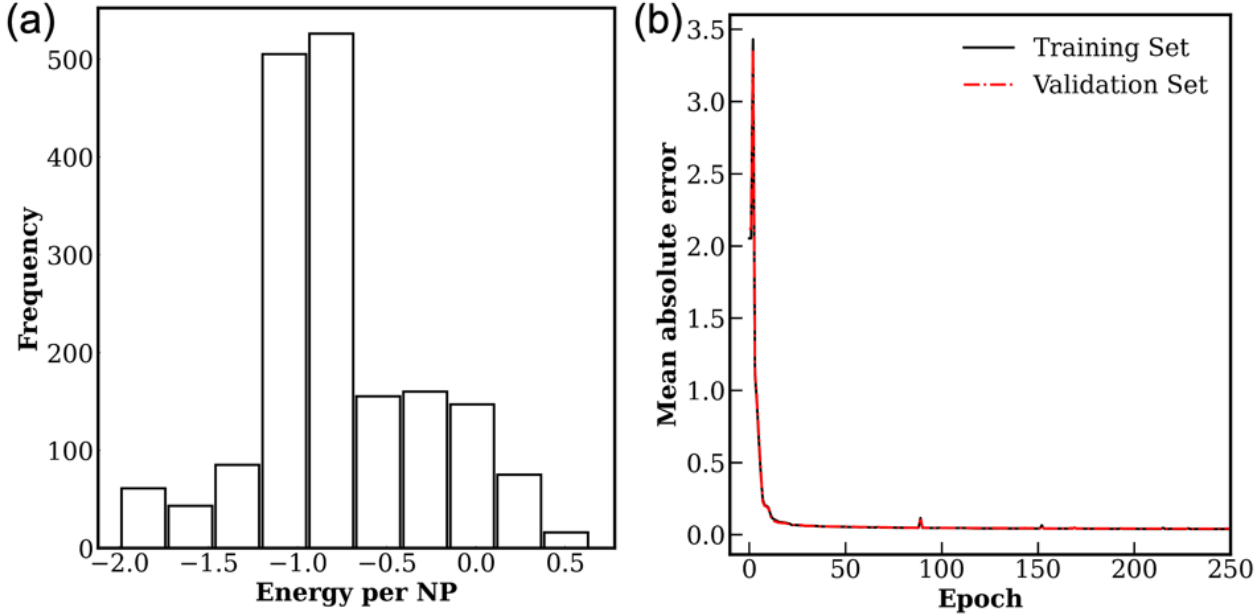

*Figure 2: The energy distribution of all the GNP clusters that are collected for DNN model development is shown in (a), and the mean absolute error during the training is shown in (b) for training and validation data set. The energies are in LJ unit. The training and validation set consist of 5600 and 800 points, respectively.*

The energy distribution of training data that are collected during the temperature quenching simulations of GNP clusters is shown in Figure 2a. These configurations are sampled within a temperature range of T*=1.5 to T*=0.2. We use 80% of these configurations for training the models, and the remaining 20% of the data is used to test the performance of the models. A validation data set is also created by randomly selecting 10% of the training data that are used for cross-validation of the models' performance during the training. As mentioned earlier, the Cartesian coordinates of the centroid of NPs and the point of grafting are converted to translationally and rotationally invariant coordinates and feed to the DNNs during the training. The mean absolute error (MAE), which is the difference between the predicted energy at the network's output (cf. Figure 1) and the actual energy of a GNP cluster during the training is plotted in Figure 2b. During the training, the hyperparameters of the DNNs are optimized to minimize the MAE. A rapid reduction of MAE is observed during the early stage of the training and MAE is less than 2% of the actual energy value within the 50 training cycles as shown in Figure 2b for both the training and validation data sets. We compare the actual and predicted energies for training and test data sets in Figure 3. The coefficient of determination ($R^2$) of the model is above 0.99 for both the training and test sets. This suggests that the DNNs can accurately capture the energy surface of a GNP cluster. As mentioned earlier, the DNNs is made of two types of DNN – one represents the local interacting environment of the centroid of an NP, and the 2nd one represents that of the point of grafting on an NP's surface. We, therefore, infer that the potential energy surface of an individual GNP can be represented by these two types of DNN.

We now conduct MD simulations of nanoparticles based on the energy predicted by the deep learning models. In these simulations, a polymer grafted nanoparticle is represented by two DNNs – one predicts the potential energy of the centroid of a nanoparticle, and the second one predicts the potential energy at the point of grafting on a nanoparticle's surface. MD simulations of NPs interacting via these DL-PMF are carried out using LAMMPS[50]. These MD simulations are performed in a bulk 3D environment in a periodic simulation box with a fixed volume fraction. The number of particles in these simulations are 100. Each particle has two interaction sites represented by the two DNNs. The effective force on an NP is calculated from the predicted energy. For a particle $k$, along a Cartesian direction $r_{k,\alpha}$, $\alpha = \{x, y, z\}$, the force can be written as[43] $F_{k\alpha} = -\frac{\partial E}{\partial r_{k\alpha}} = -\sum_{i=1}^{N_A} \frac{\partial E_i}{\partial r_{k,\alpha}} = \sum_{i=1}^{N_A} \sum_{j=1}^{M_i} \frac{\partial E_i}{\partial G_{ij}} \frac{\partial G_{ij}}{\partial r_{k\alpha}}$. Here, $N_A$ is the number of particles within the cut-off distance, and $G_{ij}$ is the $j^{th}$ symmetry function for the $i^{th}$ particle. The variable $M_i$ varies from 1 to 50, which is the total number of symmetry functions. Within this framework, we treat an NP and its patch as a rigid body. The total force and torque of the rigid body is computed as the sum of that of the NP and its patch. At each timestep of the MD simulation, the coordinates and velocities are updated so that the centroid of the NP and patch moves as a single entity.[47] The simulations are carried out in an implicit solvent condition in an NVT ensemble, wherein the temperature is controlled using the Langevin thermostat. We use the velocity-Verlet algorithm with a timestep of $0.005\tau$ to integrate the equation of motion. We conduct DL-PMF-based MD simulations of NPs for several volume fractions, which is defined as $\phi = NV_0/V$. Here, N, $V_0$ and V correspond to the number of nanoparticles in the system, the volume of a nanoparticle and the volume of the simulation box, respectively. We conduct MD simulations across a range of temperatures (T*=1.5-0.2) to determine the stability and transferability of the DL-PMF. The DL-PMF produces stable MD trajectories for all temperatures, and we observe continuous aggregations of nanoparticles upon cooling. The potential energy of the system at different temperatures during a cooling-heating cycle is shown in SI. The predicted NPs self-assembly based on DL-PMF at T*=0.2 is shown in Figure 4. We further carry out MD simulations of actual GNPs in the same condition to verify the prediction of the deep learning models, which are also shown in Figure 4. For very low volume fraction ($\phi = 0.001$), the DL-PMF predicts spherical aggregates of NPs and it is consistent with the actual single-chain grafted NPs assembly as shown in Figure 4a-c. As the NPs' volume fraction increases, we observe a percolating network-like superstructure followed by a bilayer assembly. This is consistent with their actual assembly as shown in Figure 4d-i. The DL-PMF predicted radial distribution functions for all three cases are in close agreement with the actual RDFs of the NPs. Therefore, we infer that the DL-PMF captures the anisotropic interactions in a polymer grafted NPs system very accurately and yields structural properties of the system, which are comparable with the actual superstructures.

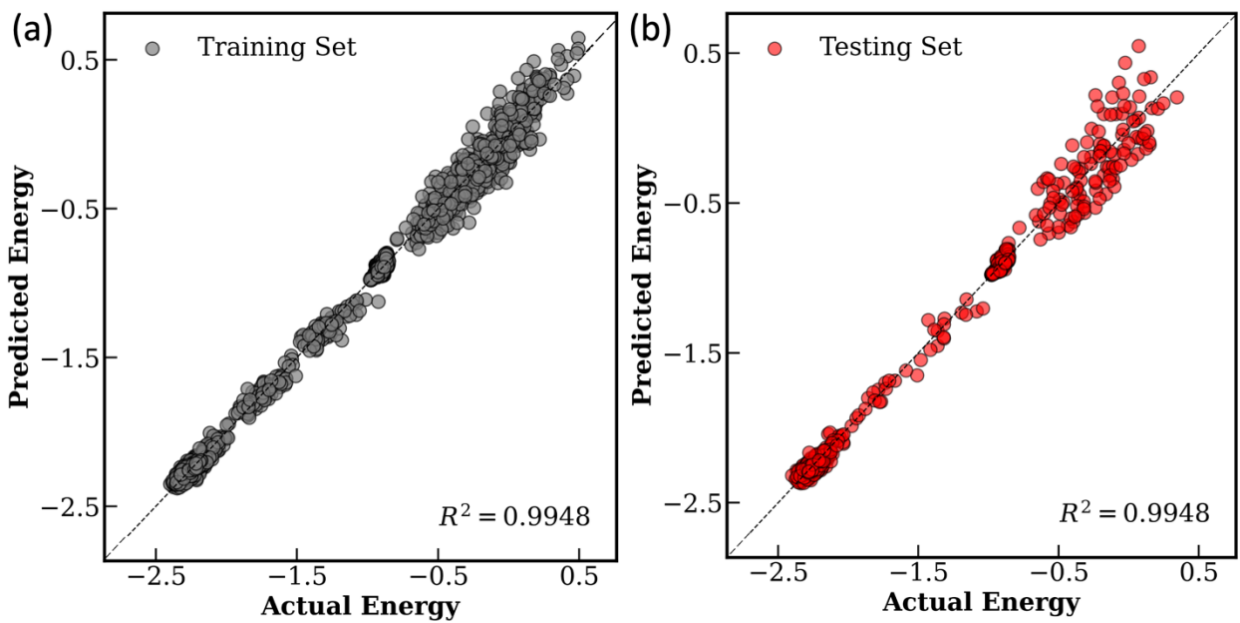

*Figure 3: Performance of the DNNs. The predicted energy is plotted against the actual energy of clusters of nanoparticles for (a) training data set and (b) test data set, respectively. The dotted lines are the x=y lines. The energies are in LJ unit and normalized by number of nanoparticles in a cluster. The training and test set consist of ~5600 and 1600 data points.*

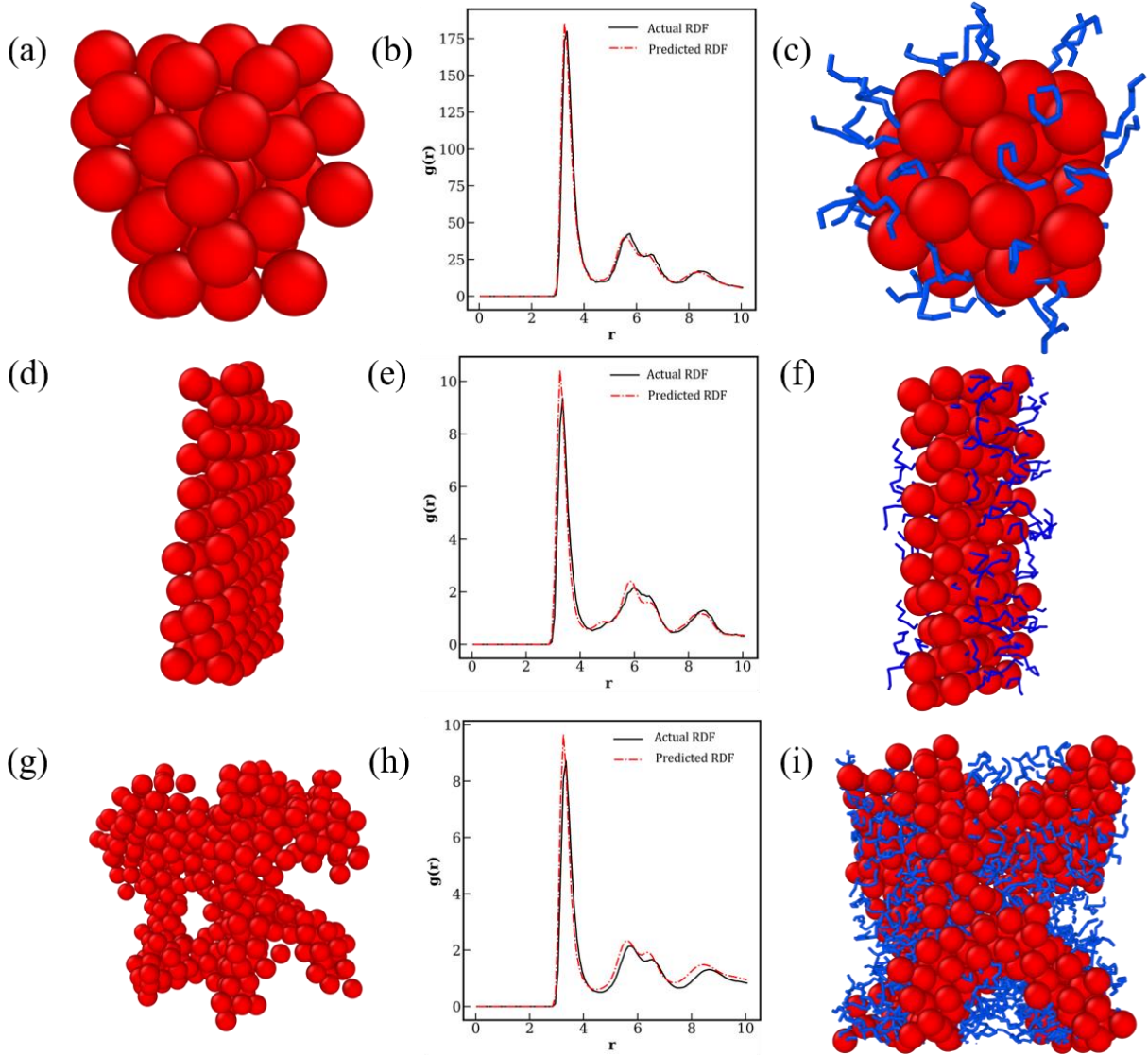

*Figure 4: The DL-PMF predicted self-assembly of polymer grafted nanoparticles. The MD snapshots of the predicted self-assembled structures are shown in the first column while the actual single chain grafted nanoparticles assembles are shown in the last column. The middle column compares the radial distribution functions for the two cases. The first row (a,b,c) corresponds to volume fraction 0.001, while the second row (d,e,f) and third row (g,h,i) correspond to volume fractions 0.15 and 0.24, respectively.*

In summary, we propose a new approach to compute the potential of mean force between a pair of polymer grafted nanoparticles using deep learning. We conduct molecular dynamics simulation based on the prediction of the deep learning model and demonstrate that the deep learning potential of mean force yields very accurate self-assembled superstructures of a large number of polymer grafted nanoparticles. We find this deep learning approach very efficient and accurate in capturing anisotropic interactions, and it predicts long-range anisotropic aggregates of polymer grafted nanoparticles. In the present study, we use a very generic phenomenological model of polymer and nanoparticle to construct a deep learning framework that predicts the potential of mean force between single-chain grafted nanoparticles. This framework can be further expanded to study effective interaction in several other nanocomposite systems, including multiple-chain grafted nanoparticles, polymer grafted nanoparticles in a polymer matrix and bare nanoparticles in a polymer matrix. Moreover, we expect that the potential of mean force for systems with atomistic details can be modeled using this deep learning framework.

## ACKNOWLEDGMENT

The work is made possible by financial support from the SERB, DST, Gov. of India through a start-up research grant (SRG/2020/001045) and the National Supercomputing Mission's research grant (DST/NSM/R&D_HPC_Applications/2021/40). This research used resources of the Argonne Leadership Computing Facility, which is a DOE Office of Science User Facility supported under Contract DE-AC02-06CH11357. We also used the computational facility of the Center for Nanoscience Materials. Use of the Center for Nanoscale Materials, an Office of Science user facility, was supported by the U.S. Department of Energy, Office of Science, Office of Basic Energy Sciences, under Contract No. DE-AC02-06CH11357. We acknowledge the use of the computing resources at HPCE, IIT Madras.## REFERENCE

(1) Israelachvili, J. N. *Intermolecular and Surface Forces*, 3rd edition.; Academic Press: Burlington (Mass.), 2011.

(2) Poon, W. C. K. The Physics of a Model Colloid Polymer Mixture. *J. Phys. Condens. Matter* **2002**, *14* (33), R859–R880. https://doi.org/10.1088/0953-8984/14/33/201.

(3) Pusey, P. N.; Zaccarelli, E.; Valeriani, C.; Sanz, E.; Poon, W. C. K.; Cates, M. E. Hard Spheres: Crystallization and Glass Formation. *Philos. Trans. R. Soc. Math. Phys. Eng. Sci.* **2009**, *367* (1909), 4993–5011. https://doi.org/10.1098/rsta.2009.0181.

(4) Zaccarelli, E.; Liddle, S. M.; Poon, W. C. K. On Polydispersity and the Hard Sphere Glass Transition. *Soft Matter* **2015**, *11* (2), 324–330. https://doi.org/10.1039/C4SM02321H.

(5) Ilett, S. M.; Orrock, A.; Poon, W. C. K.; Pusey, P. N. Phase Behavior of a Model Colloid-Polymer Mixture. *Phys. Rev. E* **1995**, *51* (2), 1344–1352. https://doi.org/10.1103/PhysRevE.51.1344.

(6) Rivas-Barbosa, R.; Ruiz-Franco, J.; Lara-Peña, M. A.; Cardellini, J.; Licea-Claverie, A.; Camerin, F.; Zaccarelli, E.; Laurati, M. Link between Morphology, Structure, and Interactions of Composite Microgels. *Macromolecules* **2022**, *55* (5), 1834–1843. https://doi.org/10.1021/acs.macromol.1c02171.

(7) Kumar, S. K.; Ganesan, V.; Riggleman, R. A. Perspective: Outstanding Theoretical Questions in Polymer-Nanoparticle Hybrids. *J. Chem. Phys.* **2017**, *147* (2), 020901. https://doi.org/10.1063/1.4990501.

(8) Cheng, S.; Stevens, M. J.; Grest, G. S. Ordering Nanoparticles with Polymer Brushes. *J. Chem. Phys.* **2017**, *147* (22), 224901. https://doi.org/10.1063/1.5006048.

(9) Asakura, S.; Oosawa, F. On Interaction between Two Bodies Immersed in a Solution of Macromolecules. *J. Chem. Phys.* **1954**, *22* (7), 1255–1256. https://doi.org/10.1063/1.1740347.

(10) Asakura, S.; Oosawa, F. Interaction between Particles Suspended in Solutions of Macromolecules. *J. Polym. Sci.* **1958**, *33* (126), 183–192. https://doi.org/10.1002/pol.1958.1203312618.

(11) Miyazaki, K.; Schweizer, K. S.; Thirumalai, D.; Tuinier, R.; Zaccarelli, E. The Asakura–Oosawa Theory: Entropic Forces in Physics, Biology, and Soft Matter. *J. Chem. Phys.* **2022**, *156* (8), 080401. https://doi.org/10.1063/5.0085965.

(12) Akcora, P.; Liu, H.; Kumar, S. K.; Moll, J.; Li, Y.; Benicewicz, B. C.; Schadler, L. S.; Acehan, D.; Panagiotopoulos, A. Z.; Pryamitsyn, V.; Ganesan, V.; Ilavsky, J.; Thiyagarajan, P.; Colby, R. H.; Douglas,

J. F. Anisotropic Self-Assembly of Spherical Polymer-Grafted Nanoparticles. *Nat. Mater.* **2009**, *8* (4), 354–359. https://doi.org/10.1038/nmat2404.

(13) Zhang; Horsch, M. A.; Lamm, M. H.; Glotzer, S. C. Tethered Nano Building Blocks: Toward a Conceptual Framework for Nanoparticle Self-Assembly. *Nano Lett.* **2003**, *3* (10), 1341–1346. https://doi.org/10.1021/nl034454g.

(14) Zhang, Z.; Glotzer, S. C. Self-Assembly of Patchy Particles. *Nano Lett.* **2004**, *4* (8), 1407–1413. https://doi.org/10.1021/nl0493500.

(15) Pothukuchi, R. P.; Prajapat, V. K.; Radhakrishna, M. Charge-Driven Self-Assembly of Polyelectrolyte-Grafted Nanoparticles in Solutions. *Langmuir* **2021**, *37* (41), 12007–12015. https://doi.org/10.1021/acs.langmuir.1c01571.

(16) Sgouros, A. P.; Revelas, C. J.; Lakkas, A. T.; Theodorou, D. N. Potential of Mean Force between Bare or Grafted Silica/Polystyrene Surfaces from Self-Consistent Field Theory. *Polymers* **2021**, *13* (8), 1197. https://doi.org/10.3390/polym13081197.

(17) Jiao, Y.; Akcora, P. Understanding the Role of Grafted Polystyrene Chain Conformation in Assembly of Magnetic Nanoparticles. *Phys. Rev. E* **2014**, *90* (4), 042601. https://doi.org/10.1103/PhysRevE.90.042601.

(18) Hooper, J. B.; Schweizer, K. S. Contact Aggregation, Bridging, and Steric Stabilization in Dense Polymer–Particle Mixtures. *Macromolecules* **2005**, *38* (21), 8858–8869. https://doi.org/10.1021/ma051318k.

(19) Hall, L. M.; Jayaraman, A.; Schweizer, K. S. Molecular Theories of Polymer Nanocomposites. *Curr. Opin. Solid State Mater. Sci.* **2010**, *14* (2), 38–48. https://doi.org/10.1016/j.cossms.2009.08.004.

(20) Jayaraman, A.; Schweizer, K. S. Effective Interactions, Structure, and Phase Behavior of Lightly Tethered Nanoparticles in Polymer Melts. *Macromolecules* **2008**, *41* (23), 9430–9438. https://doi.org/10.1021/ma801722m.

(21) Hooper, J. B.; Schweizer, K. S. Theory of Phase Separation in Polymer Nanocomposites. *Macromolecules* **2006**, *39* (15), 5133–5142. https://doi.org/10.1021/ma060577m.

(22) Hooper, J. B.; Schweizer, K. S.; Desai, T. G.; Koshy, R.; Keblinski, P. Structure, Surface Excess and Effective Interactions in Polymer Nanocomposite Melts and Concentrated Solutions. *J. Chem. Phys.* **2004**, *121* (14), 6986–6997. https://doi.org/10.1063/1.1790831.

(23) Ganesan, V.; Jayaraman, A. Theory and Simulation Studies of Effective Interactions, Phase Behavior and Morphology in Polymer Nanocomposites. *Soft Matter* **2013**, *10* (1), 13–38. https://doi.org/10.1039/C3SM51864G.

(24) Patra, T. K.; Singh, J. K. Coarse-Grain Molecular Dynamics Simulations of Nanoparticle-Polymer Melt: Dispersion vs. Agglomeration. *J. Chem. Phys.* **2013**, *138* (14), 144901. https://doi.org/10.1063/1.4799265.

(25) Liu, J.; Gao, Y.; Cao, D.; Zhang, L.; Guo, Z. Nanoparticle Dispersion and Aggregation in Polymer Nanocomposites: Insights from Molecular Dynamics Simulation. *Langmuir* **2011**, *27* (12), 7926–7933. https://doi.org/10.1021/la201073m.

(26) Handle, P. H.; Zaccarelli, E.; Gnan, N. Effective Potentials Induced by Mixtures of Patchy and Hard Co-Solutes. *J. Chem. Phys.* **2021**, *155* (6), 064901. https://doi.org/10.1063/5.0059304.

(27) Erigi, U.; Dhumal, U.; Tripathy, M. Phase Behavior of Polymer–Nanorod Composites: A Comparative Study Using PRISM Theory and Molecular Dynamics Simulations. *J. Chem. Phys.* **2021**, *154* (12), 124903. https://doi.org/10.1063/5.0038186.

(28) Patra, T. K.; Singh, J. K. Polymer Directed Aggregation and Dispersion of Anisotropic Nanoparticles. *Soft Matter* **2014**, *10* (11), 1823–1830. https://doi.org/10.1039/C3SM52216D.

(29) Lu, S.; Wu, Z.; Jayaraman, A. Molecular Modeling and Simulation of Polymer Nanocomposites with Nanorod Fillers. *J. Phys. Chem. B* **2021**, *125* (9), 2435–2449. https://doi.org/10.1021/acs.jpcb.1c00097.

(30) Gollanapalli, V.; Manthri, A.; Sankar, U. K.; Tripathy, M. Dispersion, Phase Separation, and Self-Assembly of Polymer-Grafted Nanorod Composites. *Macromolecules* **2017**, *50* (21), 8816–8826. https://doi.org/10.1021/acs.macromol.7b01754.

(31) Martin, T. B.; Jayaraman, A. Using Theory and Simulations To Calculate Effective Interactions in Polymer Nanocomposites with Polymer-Grafted Nanoparticles. *Macromolecules* **2016**, *49* (24), 9684–9692. https://doi.org/10.1021/acs.macromol.6b01920.

(32) Lu, S.; Jayaraman, A. Effect of Nanorod Physical Roughness on the Aggregation and Percolation of Nanorods in Polymer Nanocomposites. *ACS Macro Lett.* **2021**, *10* (11), 1416–1422. https://doi.org/10.1021/acsmacrolett.1c00503.

(33) Moinuddin, M.; Biswas, P.; Tripathy, M. The Effect of Surface Roughness on the Phase Behavior of Colloidal Particles. *J. Chem. Phys.* **2020**, *152* (4), 044902. https://doi.org/10.1063/1.5136080.

(34) Martin, T. B.; Jayaraman, A. Identifying the Ideal Characteristics of the Grafted Polymer Chain Length Distribution for Maximizing Dispersion of Polymer Grafted Nanoparticles in a Polymer Matrix. *Macromolecules* **2013**, *46* (22), 9144–9150. https://doi.org/10.1021/ma401763y.

(35) Kalb, J.; Dukes, D.; Kumar, S. K.; Hoy, R. S.; Grest, G. S. End Grafted Polymernanoparticles in a Polymeric Matrix: Effect of Coverage and Curvature. *Soft Matter* **2011**, *7* (4), 1418–1425. https://doi.org/10.1039/C0SM00725K.

(36) Modica, K. J.; Martin, T. B.; Jayaraman, A. Effect of Polymer Architecture on the Structure and Interactions of Polymer Grafted Particles: Theory and Simulations. *Macromolecules* **2017**, *50* (12), 4854–4866. https://doi.org/10.1021/acs.macromol.7b00524.

(37) Martin, T. B.; Jayaraman, A. Effect of Matrix Bidispersity on the Morphology of Polymer-Grafted Nanoparticle-Filled Polymer Nanocomposites. *J. Polym. Sci. Part B Polym. Phys.* **2014**, *52* (24), 1661–1668. https://doi.org/10.1002/polb.23517.

(38) Nair, N.; Wentzel, N.; Jayaraman, A. Effect of Bidispersity in Grafted Chain Length on Grafted Chain Conformations and Potential of Mean Force between Polymer Grafted Nanoparticles in a Homopolymer Matrix. *J. Chem. Phys.* **2011**, *134* (19), 194906. https://doi.org/10.1063/1.3590275.

(39) Martin, T. B.; Dodd, P. M.; Jayaraman, A. Polydispersity for Tuning the Potential of Mean Force between Polymer Grafted Nanoparticles in a Polymer Matrix. *Phys. Rev. Lett.* **2013**, *110* (1), 018301. https://doi.org/10.1103/PhysRevLett.110.018301.

(40) Santos, A.; Singh, C.; Glotzer, S. C. Coarse-Grained Models of Tethers for Fast Self-Assembly Simulations. *Phys. Rev. E* **2010**, *81* (1), 011113. https://doi.org/10.1103/PhysRevE.81.011113.

(41) Bozorgui, B.; Meng, D.; Kumar, S. K.; Chakravarty, C.; Cacciuto, A. Fluctuation-Driven Anisotropic Assembly in Nanoscale Systems. *Nano Lett.* **2013**, *13* (6), 2732–2737. https://doi.org/10.1021/nl401378r.

(42) Behler, J.; Parrinello, M. Generalized Neural-Network Representation of High-Dimensional Potential-Energy Surfaces. *Phys. Rev. Lett.* **2007**, *98* (14), 146401. https://doi.org/10.1103/PhysRevLett.98.146401.

(43) Behler, J. Atom-Centered Symmetry Functions for Constructing High-Dimensional Neural Network Potentials. *J. Chem. Phys.* **2011**, *134* (7), 074106. https://doi.org/10.1063/1.3553717.

(44) Iacovella, C. R.; Keys, A. S.; Horsch, M. A.; Glotzer, S. C. Icosahedral Packing of Polymer-Tethered Nanospheres and Stabilization of the Gyroid Phase. *Phys. Rev. E* **2007**, *75* (4), 040801. https://doi.org/10.1103/PhysRevE.75.040801.

(45) Phillips, C. L.; Iacovella, C. R.; Glotzer, S. C. Stability of the Double Gyroid Phase to Nanoparticle Polydispersity in Polymer-Tethered Nanosphere Systems. *Soft Matter* **2010**, *6* (8), 1693–1703. https://doi.org/10.1039/B911140A.

(46) Kremer, K.; Grest, G. S. Dynamics of Entangled Linear Polymer Melts: A Molecular-dynamics Simulation. *J. Chem. Phys.* **1990**, *92* (8), 5057–5086. https://doi.org/10.1063/1.458541.

(47) Miller, T. F.; Eleftheriou, M.; Pattnaik, P.; Ndirango, A.; Newns, D.; Martyna, G. J. Symplectic Quaternion Scheme for Biophysical Molecular Dynamics. *J. Chem. Phys.* **2002**, *116* (20), 8649–8659. https://doi.org/10.1063/1.1473654.

(48) Plimpton, S. Fast Parallel Algorithms for Short-Range Molecular Dynamics. *J. Comput. Phys.* **1995**, *117* (1), 1–19. https://doi.org/10.1006/jcph.1995.1039.

(49) Artrith, N.; Urban, A. An Implementation of Artificial Neural-Network Potentials for Atomistic Materials Simulations: Performance for TiO2. *Comput. Mater. Sci.* **2016**, *114*, 135–150. https://doi.org/10.1016/j.commatsci.2015.11.047.

(50) *LAMMPS Molecular Dynamics Simulator*. https://www.lammps.org/ (accessed 2021-09-13).